\documentclass[a4paper,fleqn,usenatbib]{mn2e}
\usepackage{amssymb}
\usepackage{textcomp}
\usepackage{graphicx}
\usepackage{txfonts}
\usepackage{epstopdf} 
\usepackage{xcolor}     % For text colors
\begin{document}

\title[Follow-up of (6478) Gault]{Physical characterization of the active asteroid 
(6478) Gault\thanks{Based on observations collected at the Cassini Telescope of the Loiano Observatory, Italy}}
\author[Carbognani, Buzzoni, Stirpe]{Albino Carbognani\thanks{E-mail: albino.carbognani@inaf.it}
Alberto Buzzoni, Giovanna Stirpe\\
INAF - Osservatorio di Astrofisica e Scienza dello Spazio, Via Gobetti 93/3, 40129 Bologna, Italy
}

%\offprints{}

\date{Received ; Accepted}

\maketitle

\label{firstpage}

\begin{abstract}
We report dense lightcurve photometry, $BVR_{c}$ colors and phase - mag curve of (6478) Gault, an active asteroid with sporadic comet-like ejection of dust. We collected optical observations along the 2020 Jul-Nov months during which the asteroid appear always star-like, without any form of perceptible activity. We found complex lightcurves, with low amplitude around opposition and a bit higher amplitude far opposition, with a mean best rotation period of $2.46_{\pm 0.02}$ h. Shape changes were observed in the phased lightcurves after opposition, a probable indication of concavities and surface irregularities. We suspect the existence of an Amplitude-Phase Relationship in $C$ band. The mean colors are $B-V = +0.84_{\pm 0.04}$, $V-R_{c} = +0.43_{\pm 0.03}$ and 
$B-R_{c} = +1.27_{\pm 0.02}$, compatible with an S-type asteroid, but variables with the rotational phase index of a non-homogeneous surface composition. From our phase - mag curve and Shevchenko's empirical photometric system, the geometric albedo result $p_V=0.13_{\pm 0.04}$, lower than the average value of the S-class. We estimate an absolute mag in $V$ band of about +14.9 and this, together with the albedo value, allows to estimate a diameter of about 3-4 km, so Gault may be smaller than previously thought.

\end{abstract}


%%%%%%%%%%%%%%%%%%%%%%%%%%%%%%%%%%%%%%%%%%%%%%%%%%%%%%%%%%%%%%%%%%%%%%%%%%
\section{Introduction}
\label{sec:introduction}

Main belt asteroid (6478) Gault (hereafter ``Gault''), two years ago surged to very special attention as an outstanding member of the active asteroids class, sporting typical morphological features of comets, such as a coma and tail \citep{smith2019, jewitt2019, kleyna2019, quanzhi2019}. Gault was not the first asteroid that developed the phenomenological traits typical of a comet, almost 40 are now known. Previously they were known as main-belt comets, but the class has been changed in active asteroids because the causes of the activity can be heterogeneous: sublimation of volatile materials, rotational disintegration, thermal fracturing or collision with smaller asteroids \citep{jewitt2012}. Among the most famous members of this class there is the asteroid (7968) Elst-Pizarro (also known as comet 133P/Elst-Pizarro).\\
In a previous work, hereafter Paper I \citep{carbognani2020}, we had established that Gault's rotation period was between 3.3 and 3.4 h and that the color indices changed with the rotation phase, i.e. at least a part of the surface was bluer than the rest. In this paper we present further physical observations conducted on Gault during a quiescence phase in 2020.
 
\section{Observations \& Data reduction}
\label{sec:instruments}
We surveyed Gault along the 2020 Jul-Nov period, with the aim of physically characterizing the asteroid with optical observations from ground. During this period, Gault's opposition occurred on Sep 28, 2020, reaching a minimum phase angle of about 0.46 degrees.\\
We have used the ``G.D. Cassini'' 152~cm f/4.6 Ritchey-Chr\'etien telescope of the Loiano Observatory (Bologna, Italy, IAU 598). The BFOSC camera was attached the telescope, equipped with a Princeton Instruments EEV $1340 \times 1300$ pixel back-illuminated CCD with 20~$\mu$m pixel size. Platescale in bin 2 mode was 1.16~arcsec~px$^{-1}$ leading to a field of view of $13.0 \times 12.6$~arcmin. Broad-band Johnson/Cousins $B,V,R_{c}$ filters were used to measure asteroid's colors. For dense photometry the clear filter (hereafter $C$) was used, in order to maximize the signal to noise ratio (S/N), and the telescope was tracked at non-sidereal rates to follow Gault's motion and further increase S/N. Images standard processing included bias subtraction and flat fielding procedure. \\
We collected about 45 h of observation on the target, as summarized in Table~\ref{t01}. In all the sessions the sky was clear - the only exceptions are the sessions of 12 and 18 Oct - and without moonlight. Regarding the colors $B, V, R_{c}$ measurements, the images were taken during nights with stable transparency conditions. The photometric reduction has been carried out according to the usual standard calibration procedure \citep{landolt1992, harris1981}. In addition, special care has been devoted to take the Landolt field at similar airmass than Gault's frames in order to minimize differential corrections, as in Paper I.\\

\begin{table*}
\centering
\caption{Summary of the Gault's 2020 observing sessions, both dense photometry with clear filter ($C$) and $BVR_{c}$ for colors determination. In the FWHM (Full Width Half Maximum) column, the Gault value is shown (measured by averaging over all the images taken for dense photometry), while in round brackets there is a comparison with the FWHM of the field stars: no significant differences appear (the FWHM uncertainty is of the order of 0.1-0.2 arcsec). The FWHM of the stars can be taken also as a local seeing estimate.}
\label{t01}

\begin{tabular}{lccccccc}
\hline
Mean Date    & Dense photometry & No. of   & Exposure & Timespan & FWHM     & Colors    & Sky conditions\\
             & band             & frames   & [s]      & [h]      & [arcsec] &           & during session  \\
\hline                 
July 21 & $R_{c}$          & 26 &  240 & 1.7 & 2.5 (stars 2.5)  & ---       & Clear\\
July 22 & $R_{c}$          & 24 &  240 & 1.5 & 2.3 (stars 2.3)  & ---       & Clear\\
July 24 & $R_{c}$          & 26 &  300 & 2.2 & 2.5 (stars 2.5)  & ---       & Clear\\
July 27 & $R_{c}$          & 25 &  240 & 1.7 & 3.8 (stars 3.5)  & ---       & Clear\\
July 29 & $C$              & 29 &  240 & 2.0 & 2.9 (stars 2.9)  & ---       & Clear - short\\
July 31 & $C$              & 31 &  240 & 1.7 & 2.8 (stars 3.1)  & ---       & Clear - short\\
Aug 22  & $C$              & 76 &  180 & 3.7 & 3.0 (stars 2.5)  & ---       & Clear\\
Sep 15  & $C$              & 68 &  180 & 3.8 & 3.0 (stars 2.7)  & ---       & Clear\\
Sep 17  & $C$              & 82 &  180 & 4.2 & 3.5 (stars 3.5)  & $BVR_{c}$ & Clear\\
Oct 12  & $C$              & 26 &  180 & 1.0 & 3.8 (stars 3.8)  & ---       & Cirrus - rejected\\
Oct 13  & $C$              & 57 &  180 & 2.6 & 3.2 (stars 3.4)  & $BVR_{c}$ & Clear - short\\
Oct 16  & ---              &--- &  --- & --- & ---              & $BVR_{c}$ & Clear\\
Oct 18  & ---              &--- &  --- & --- & ---              & $BVR_{c}$ & Cirrus - rejected\\
Oct 20  & $C$              & 66 &  180 & 3.0 & 3.1 (stars 3.4)  & $BVR_{c}$ & Clear\\
Oct 21  & $C$              & 76 &  180 & 4.0 & 3.9 (stars 3.9)  & ---       & Clear - bad seeing, rejected\\
Nov 07  & $C$              & 35 &  180 & 1.8 & 2.7 (stars 2.5)  & $BVR_{c}$ & Clear - short\\

\hline
\end{tabular}
\end{table*}

\section{Rotation period}
\label{sec:lightcurves}
Determining Gault's rotation period was a much more complex task than expected, even with the asteroid's activity reduced to zero (see section~\ref{sec:colors}). MPO {\sc Canopus} package \citep{warner2009} was used for differential aperture photometry and period determination with the classical FALC algorithm \citep{harris1989}. {\sc Peranso 3} package\footnote{http://www.peranso.com} was also used for an independent analysis of the period with the Fourier-component Analysis of Variance, i.e. ANOVA algorithm \citep{schwa1996}. Given the peculiarity of Gault we kept the lightcurves from too different sessions separate to account for changes in lighting conditions, so we decided to analyze them individually.

\subsection{July and August sessions}
\begin{figure}
\hspace*{-0.8cm}\includegraphics[width=1.1\hsize]{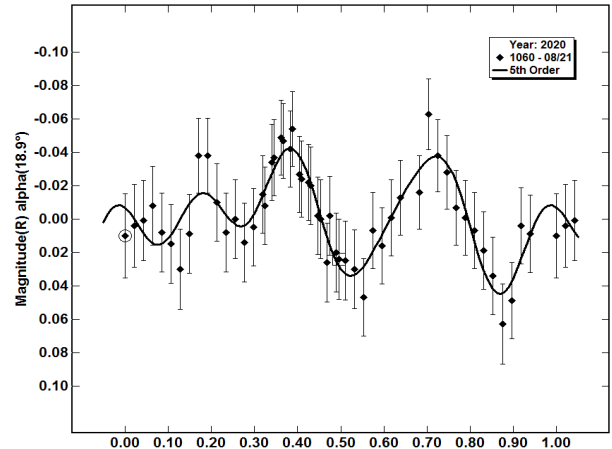}
\caption{The phased lightcurves of Aug 22 session plotted with MPO Canopus. The best period is $2.49_{\pm 0.02}$ h, with an amplitude of about 0.09 mag.
In the raw photometric data the Aug session has a gap of about 1 hour near the end, because the asteroid passed close to a very bright background star.}
\label{Gault_Aug2020}
\end{figure}

\begin{figure}
\hspace*{-0.6cm}\includegraphics[width=1.1\hsize]{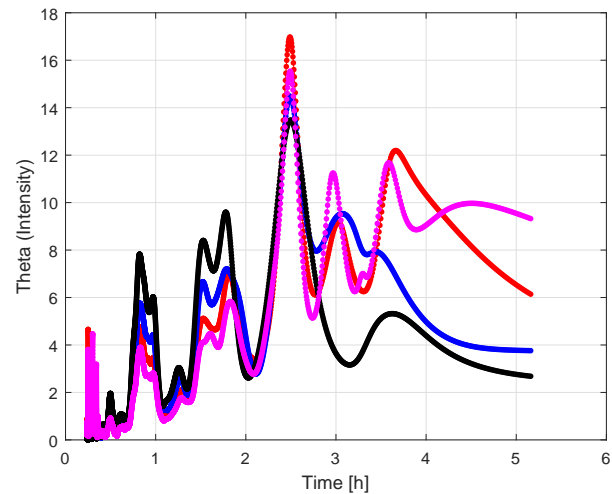}
\caption{The period spectrum of Aug 22 session according to ANOVA algorithm. The power spectrum with three harmonics is the black curve, four harmonics blue curve, five harmonics red curve and six harmonics is the magenta curve. The maximum peak is around 2.49 h.}
\label{Gault_PS_Aug2020}
\end{figure}

The Jul sessions are short (see Table~\ref{t01}) and with too low S/N to be useful for period analysis. They are useful for monitoring asteroid activity and phase-magnitude curve determination instead (see sections~\ref{sec:colors} and ~\ref{sec:phasecurve}). \\
The Aug session is good as S/N, the lightcurve has a low amplitude ($0.09_{\pm 0.02}$ mag), with 4 maxima and minima, very different from a typical bimodal asteroid lightcurve (Fig.~\ref{Gault_Aug2020}). Usually asteroid lightcurves are dominated by the second harmonic of the rotation period, caused by elongated shape, that give two maxima and two minima. However, if the shape is not very elongate or if the asteroid's orientation is pole-on, other harmonics may dominate so complex lightcurves are not uncommon between asteroids \citep{harris2014}. \\
The best period - both with ANOVA and FALC - is $2.49_{\pm 0.02}$ h, see Fig.~\ref{Gault_PS_Aug2020}, while the session is 3.7 h long, more than enough for lightcurve replication. 
In Fig.~\ref{Gault_PS_Aug2020} we have drawn different power spectrum using three harmonics (black curve), four harmonics (blue curve), five harmonics (red curve) and six harmonics (magenta curve). As the number of harmonics increases, the main peak remains the same while the secondary peaks, with a shorter period, tend to decrease and stabilize between five and six harmonics, an expected behavior for the alias periods \citep{schwa1996}. For this reason we have chosen to use ANOVA spectra with five harmonics.\\
The qualitative lightcurve shape and the period value are in good agreement with that found by \cite{luu2021}, who observed Gault in a single session of about 5 hours long on Aug 27, 2020 and found a best period of $2.55_{\pm 0.1}$ h. Also \cite{purdum2021}, averaging the Gault lightcurves obtained between Aug-Oct 2020, obtain a best period value of about 2.5 hours while \cite{devogele2021}, using six complete lightcurves obtained in the period Sep-Oct 2020, find $2.4929_{\pm 0.0003}$ hours. \\

\subsection{September sessions}
\begin{figure}
\hspace*{-0.8cm}\includegraphics[width=1.1\hsize]{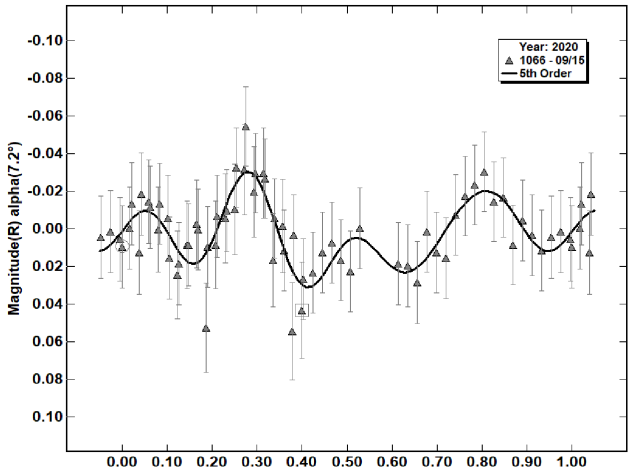}
\caption{The phased lightcurves of Sep 15 sessions plotted with MPO Canopus. The best period is $2.47_{\pm 0.02}$ h.}
\label{Gault_202009_15}
\end{figure}

\begin{figure}
\hspace*{-0.6cm}\includegraphics[width=1.1\hsize]{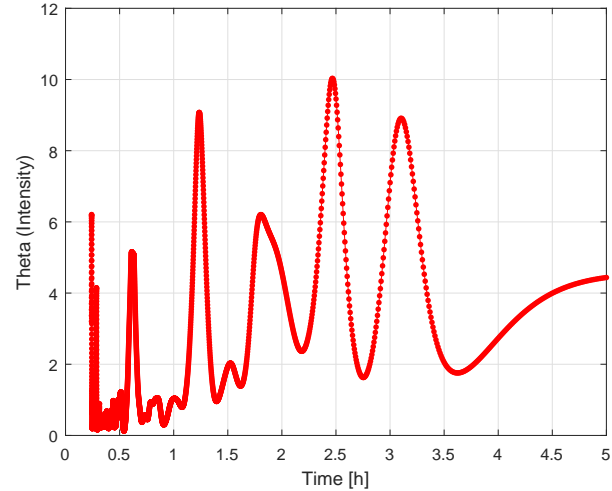}
\caption{The period spectrum of Sep 15 session according to ANOVA algorithm. }
\label{Gault_PS_202009_15}
\end{figure}

Lightcurves are taken in Sep 15 and 17, about 10 days before opposition. The Sep 15th session is a beautiful session, with very constant comparison stars, no haze, constant seeing and a long duration of about 3.8 hours. The lightcurve appears similar to the Aug session, with low amplitude (about 0.06 mag) and 4 maximum and minimum. The best period is $P = 2.47_{\pm 0.02}$ h both with ANOVA and FALC. With ANOVA also a secondary period near 3.1 h appear (see Fig.~\ref{Gault_202009_15} and Fig.~\ref{Gault_PS_202009_15}). This secondary period, also present in the Ago 22 session, is very close to the 3.34 h value we found in Paper I with the lightcurves taken in March 2019. The main peak of the period $P$ has also the peak $P/2$ (about 1.23 h, bimodal solution) and $P/4$ (about 0.62 h, monomodal solution). The secondary period at 3.1 h appear to be $5P/4$ (six-modal solution), and there is also a peak near 1.8 h or $3P/4$ (three-modal solution). Re-analyzing with ANOVA the lightcurve of Apr 15, 2019, the freest from Gault's dust emission (see Paper I, Fig. 5), the main period is approximately 3.2 hours (on a session 3.7 hour long), with a secondary period of approximately 2.4 hours: so we find period values very close to the sessions of Aug 22 and Sep 15, 2020, but with reversed intensity. For some reason that is unclear at the moment, it seems that Gault's residual activity in 2019 sessions has shown us an alias of the best period.\\
The session of Sep 17 is slightly noisier respect to previous because the field of view was more crowded with background stars and the weather conditions were also of a bit lower quality (worst seeing). Images in which the asteroid passed near faint background stars were eliminated from photometry to not contaminate the data. From the sessions of 17 Sept, Gault's lightcurve appear similar to previous session with a best period, given by FALC and ANOVA, of $2.45_{\pm 0.05}$ h, combining the two lightcurves together results a best rotation period of $2.497_{\pm 0.002}$ h.

\subsection{October and November sessions}
Lightcurves are taken in Oct 12, 13, 20 and 21, after Gault's opposition. The lightcurve of Oct 12 is short and noisy due to cirrus clouds, so was rejected. The session of Oct 13 is better but it is not long enough - 2.6 h only - to allow a sure lightcurve replication.\\ 
The session on Oct 20 is among the best, quite long (about 3.0 h), and done in good weather conditions. The lightcurve was obtained by eliminating images where Gault passes near 
background stars, in particular the last 5 images (length of about 15 minutes) were deleted. With FALC, the best period is $P = 2.43_{\pm 0.1}$ h (ANOVA give $2.42$ h, almost identical). The amplitude was rising to about $0.10_{\pm 0.02}$ mag compared to the Sep sessions and the lightcurve shows 3 maxima and 3 minima, a configuration a bit different from the pre-opposition sessions. In \cite{purdum2021} and \cite{devogele2021}, who obtained Gault's lightcurves around the same months, there is no indication of a change in the shape of the lightcurve. This change could be due to the observation with the $C$ filter instead of the $R$ filter, as we will discuss in sections~\ref{subsec:amp-phase}. \\
From ANOVA period spectrum (see Fig.~\ref{Gault_PS_20201020_ANOVA}), a peak with a period of about 3.1 hours is visible, as in Aug 22 and Sep 15 session. The lightcurve of Oct 21 session - about 4 h long - appears a bit degraded by bad seeing and has not been considered. Finally, the Nov 7 session was too short to allow an independent period to be determined, but the Gault amplitude reached the value of at least $0.15_{\pm 0.03}$ mag, helpful for amplitude-phase study (see section \ref{subsec:amp-phase}). The amplitude value was initially greater, but it was a spurious effect due to a faint background star.\\
We searched also in the Transiting Exoplanet Survey Satellite (TESS) database\footnote{https://archive.konkoly.hu/pub/tssys/dr1/} with the hope to finding Gault's photometric observations \citep{Pal2020}. According to the lightcurve derived from TESS\footnote{https://archive.konkoly.hu/pub/tssys/dr1/object\_plots/}, Gault would have a rotation period of 10.37 h with an amplitude of 0.21 mag. However the observations were made from February 28 to March 2, 2019, when Gault was still in full activity and this period value cannot reflect the rotational state of the asteroid because it was surrounded by the dust coma and still showing its tail (see Fig. 1 in Paper I).\\

\begin{figure}
\hspace*{-0.8cm}\includegraphics[width=1.1\hsize]{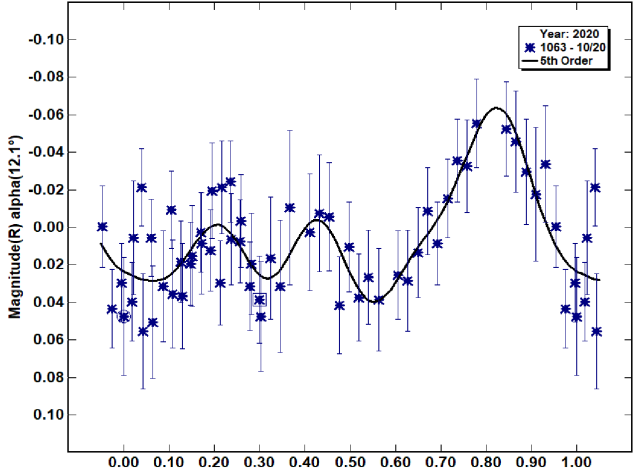}
\caption{The phased lightcurves of Oct 20 session with the best period of $2.43_{\pm 0.1}$ h.}
\label{Gault_20201020}
\end{figure}

\begin{figure}
\hspace*{-0.6cm}\includegraphics[width=1.1\hsize]{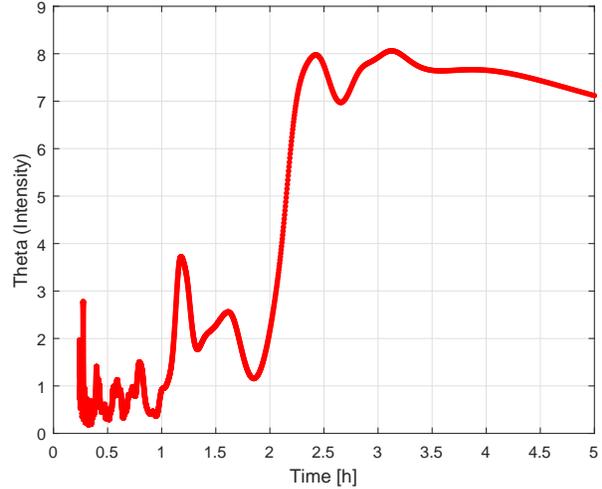}
\caption{The period spectrum of Oct 20 session according with ANOVA algorithm.}
\label{Gault_PS_20201020_ANOVA}
\end{figure}

\subsection{Conclusions about the rotation period}
The lightcurve change from Sep to Oct remember us a similar behavior of asteroid (234) Barbara. Barbara is an asteroid belonging to the inner Main Belt with a complex lightcurve, classified for a long time as a S-type. Barbara's non-convex shape was determined using both photometric and stellar occultations data \citep{tanga2015}. The shape of this asteroid is highly irregular with the presence of large concavities which cause sudden changes in the lightcurve, even for small variations in the phase angle. So Gault could be a similar case.\\
To conclude this section, taking the mean values of all the best periods (both FALC and ANOVA results), our best estimate for the Gault mean rotation period is $2.46_{\pm 0.02}$ h. This value is in good agreement with the authors already cited. In longer sessions of good quality we did not find the period found in Paper I, i.e. $3.34_{\pm 0.02}$ h, but a secondary period near 3.1 h that appear as an alias of the best period. Probably the dust that still surrounded Gault in March-April 2019 altered our photometry. Incidentally, it is interesting to note that also \cite{ivanova2020} found a secondary period of about 3.4 h with photometric observation made on February 6 and March 28, 2019.\\ 
Assuming that Gault's activity is due to expulsion of dust and stones into space due to spin-barrier limit \citep{pravec2002}, a period of $2.46_{\pm 0.02}$ hours implies a bulk density $\rho\approx 1.80_{\pm 0.03}~\textrm{g}/\textrm{cm}^3$, a value lower than the average density for S-type asteroids, about $2.72_{\pm 0.54}~\textrm{g}/\textrm{cm}^3$, see Paper I.\\
Assuming that, as S-type asteroid, Gault has the same composition as ordinary chondrites the grain densities range from $\rho_g\approx 3.75$ to $3.56~\textrm{g}/\textrm{cm}^3$ \citep{britt2001}, so the porosity $n$ of the asteroid, i.e. the fraction of empty volume with respect to the total volume, is about $n=1-\rho/\rho_g\approx 0.5$, i.e. 50\%.\\

\subsection{The amplitude-phase relationship}
\label{subsec:amp-phase}
\begin{figure}
\hspace*{-0.8cm}\includegraphics[width=1.1\hsize]{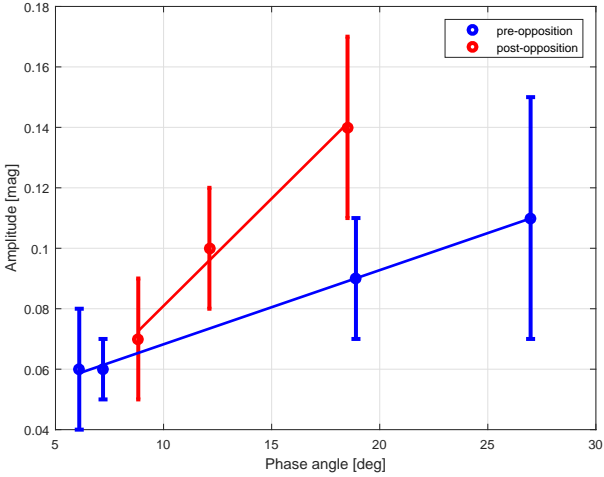}
\caption{The amplitude-phase relationship for Gault. The lower plot refers to the pre-opposition, the upper one to the post-opposition period.}
\label{fig:amp}       
\end{figure}

From the photometric observations we had not only the rotation period but also the maximum lightcurve amplitude (see Fig.~\ref{fig:amp}). While the rotation period is a fixed parameter, the amplitude depends both on the object's shape and on the aspect angle at the time of observations. However, the amplitude can be affected also by the phase angle $\alpha$ and for asteroids there is an APR (Amplitude-Phase Relationship, \cite{zappala90}). The APR can be fit by a linear equation of the form:

\begin{equation}
A\left(\alpha\right) = A\left(0^{\circ}\right)\left(1+ m\alpha\right)
\label{APR}
\end{equation}

\noindent In Eq. (\ref{APR}), $A\left(\alpha\right)$ is the lightcurve amplitude (in mag) at the phase angle $\alpha$ (in deg), while $m$ is a constant with dimension 1/deg that depends on the taxonomic type, i.e. $m = 0.030$ for S-type, $m = 0.015$ for C-type and $m = 0.013$ for M-type \citep{zappala90}. \\
In this case we also select the almost complete lightcurves of 24th Jul, 13th Oct and 7th Nov 2020 (see Table~\ref{t03}) and considered as amplitude the difference between the maximum and minimum brightness of the best fit of the phased curves. As amplitude uncertainty we have assumed the mean value of the uncertainty of the single points of the lightcurves. The measurements, given the low amplitude of the Gault's lightcurve, are delicate but from Figure \ref{fig:amp} there seems to be an APR relation. Note that there are two different plots: the lower concerns the pre-opposition and the upper one the post-opposition period. A linear fit with the least square method of Eq. \ref{APR} gives, for the lower plot: $A\left(0^{\circ}\right) = 0.04_{\pm 0.02}$ mag, compatible with the observed minimum amplitude, and $m = 0.06_{\pm 0.04}\, \hbox{deg}^{-1}$. This last value is roughly coherent with an S-type asteroid. A linear fit of the upper plot gives: $A\left(0^{\circ}\right) = 0.01_{\pm 0.05}$ mag, and $m = 0.7_{\pm 3}\, \hbox{deg}^{-1}$. In this last case the $A\left(0^{\circ}\right)$ and $m$ uncertainty are too large to be useful, i.e. they cannot be determined from the present set of measurements. These different results between the two plots are due to the fact that the pre-opposition plot has the points in a greater phase-angle range than post-opposition. \\
In \cite{devogele2021} different Gault lightcurves are shown in $R$ and $V$ band, taken in the period from 2020 Sep 11 to Oct 24, and the amplitudes appear substantially constant and equal to about 0.06 mag. In \cite{luu2021} there is only one lightcurve taken in 2020 Aug 27 with $R$ filter with an amplitude indicated by the authors equal to about 0.05 mag. In \cite{purdum2021} the average of a dozen lightcurves taken between 2020 Aug 23 and Oct 20 is shown, with an estimated amplitude of about 0.05 mag in $R$ band. From these results it can be said that in the period between 2020 Aug-Oct the Gault amplitude in $R$ band was about 0.05-0.06 mag. \\
A difference between these and our dataset is the spectral response: in order to maximize S/N we take the most complete lightcurves in $C$ band (the only lightcurve not too short taken in R is that of July 24), so we have collected also the blue part of the spectra. We suspect that observation even in the blue part of the spectrum may increase the amplitude: it is sufficient that a part of the asteroid reflects more in blue than in red and that this part, due to rotation, go in shadow to increases the lightcurve amplitude. This could provide a trail for further investigation to establish if the Gault's amplitude lightcurve depends by the filters.

\begin{table}
\centering
\caption{Gault's amplitude - phase data plotted in Fig.~\ref{fig:amp}. Pre-opposition values are with negative phase angle.}
\label{t03}

\begin{tabular}{lccc}
\hline
Mean date &  Phase ($^{\circ}$) & Amplitude (mag) & Band\\

\hline
Jul 24 & -26.98 & $0.11 \pm 0.04$ & $R_c$\\
Aug 22 & -18.90 & $0.09 \pm 0.02$ & $C$  \\
Sep 15 & -07.20 & $0.06 \pm 0.01$ & $C$  \\
Sep 17 & -06.10 & $0.06 \pm 0.02$ & $C$  \\
Oct 13 & +08.83 & $0.07 \pm 0.02$ & $C$  \\
Oct 20 & +12.12 & $0.10 \pm 0.02$ & $C$  \\
Nov 07 & +18.51 & $0.15 \pm 0.03$ & $C$  \\

\hline
\end{tabular}
\end{table}

\section{Gault's cometary activity and colors}
\label{sec:colors}

\begin{figure}
\centering
\includegraphics[width=0.8\hsize]{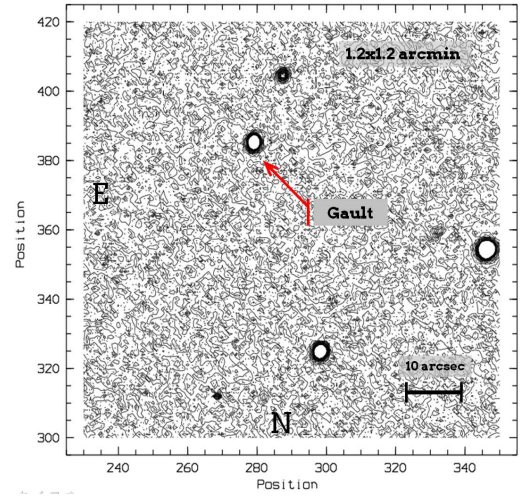}
\caption{$C$-band isophotal contour plot of Gault's images along the night of 2020 September 14-15, with 180 s exposure time. The displayed field of view is about 70 arcsec across, with North down and East to the left, as labelled. Coordinate axes are labelled in pixel scale (1 px = 0.58 arcsec). Gault is the slightly ``elongated'' object indicated by the red arrow. Gault apparent magnitude is about +17.4, while sky brightness result $\mu \sim +18 ~\hbox{mag}/\hbox{arcsec}^2$. From the images, we can rule out at a S/N = 3 confidence level, any activity signature around the asteroid brighter than $\mu \sim +22 ~\hbox{mag}/\hbox{arcsec}^2$.}
\label{Gault_back}
\end{figure}

After each dense photometric session we stacked all the calibrated images to search for a possible faint form of cometary activity. We obtained two types of mean images for each session: those stacked on Gault's angular motion and those stacked on the background stars. Comparing the FWHM of the asteroid with that of the background stars with approximately the same apparent mag of the asteroid, we found that Gault appear always star-like, without significant deviations (see Table~\ref{t01}). We also looked directly in the individual images for the presence of some form of activity with no positive results, see for example Fig.~\ref{Gault_back}. Based on these data we can say that, during our sessions, we observed the ``naked'' asteroid.\\

\begin{figure}
\hspace*{-0.8cm}\includegraphics[width=1.1\hsize]{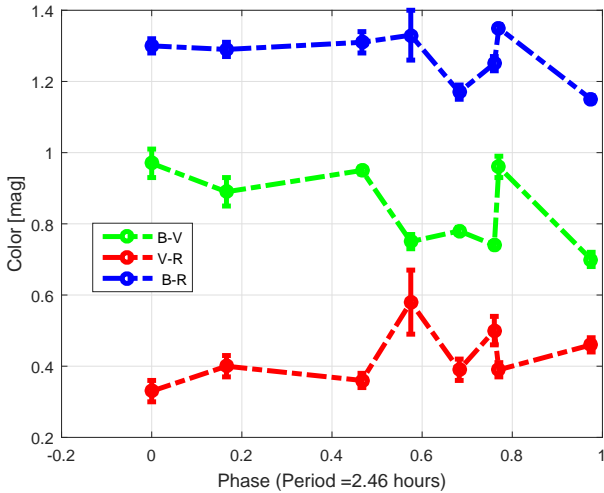}
\caption{Gault's colors vs rotational phase with 2.46 h rotation period (light-time corrected).}
\label{Gault_colors2}
\end{figure}

We have reduced\footnote{Standard calibration procedure \citep{landolt1992, harris1981}.} all the Gault's $BVR_c$ images (Table~\ref{t01}), in order to compute colors indices finding these mean values: $B-V = 0.84_{\pm 0.04}$, $V-R = 0.43_{\pm 0.03}$ and $B-R = 1.27_{\pm 0.02}$ which make it very similar to an S-type asteroid ($B-V = 0.85$, $V-R = 0.47$, $B-R = 1.32$, \cite{dandy2003}). To verify if the colors changes according to the rotational phase, we plotted the single session values (light-time corrected) with our best period. Of course the aspect angle changed during the observation months, so we haven't always seen the same asteroid's surface and the plot is not a ``map'' of the Gault's surface colors. In any case the plot is useful to see if there are macroscopic changes in colors wherever they are on surface. The results are shown in Fig.~\ref{Gault_colors2}: it appears that colors on Gault's surface change significantly, in qualitative agreement with that we found in Paper I, although we have not found colors as blue as in 2019. However in the second half of 2020 the Gault's heliocentric ecliptic longitude ranged from $345^{\circ}$ to $20^{\circ}$, while in the first half of 2019 varied in the range $160-175^{\circ}$. So in 2020 probably we observed a part of the asteroid's surface that was not visible in 2019, because the difference between the two heliocentric ecliptic longitude is near $180^{\circ}$. 

\section{Gault's mag - phase curve and effective diameter estimate}
\label{sec:phasecurve}
\begin{figure}
\hspace*{-0.8cm}\includegraphics[width=1.1\hsize]{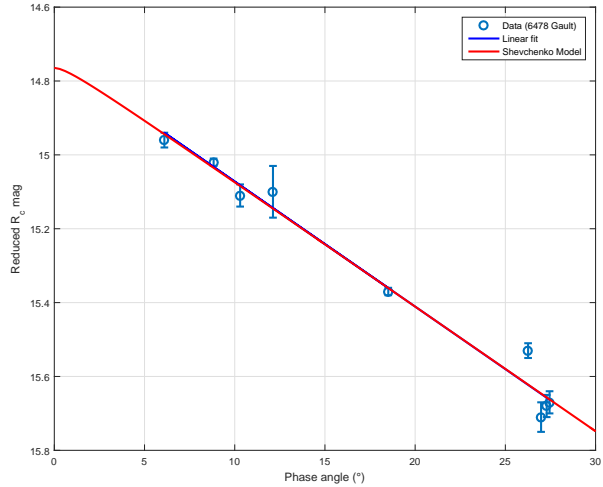}
\caption{The Gault's mag - phase curve, i.e. the $R_c$ reduced mag vs phase angle, and the best fit curve with the Shevchenko empirical system compared with linear fit.}
\label{Gault_HG}
\end{figure}

With the collected photometric data it was possible to determine the Gault phase - mag curve especially in the linear part, see Table~\ref{t02} and Fig.~\ref{Gault_HG}. In order to increase the number of photometric observations available Gaia's DR2 was consulted also, but without finding sparse Gault's observations\footnote{https://gea.esac.esa.int/archive/}. The phase - mag curve of an asteroid describes the variation of brightness, expressed in magnitudes and normalized to unit distance from Sun and observer, as a function of varying phase angle \footnote{The phase angle is the angle between the directions to the observer and to the Sun as seen from the observed body.}. \\
It is well known that the magnitude (i.e., the brightness expressed in a logarithmic scale) of small bodies of the Solar system tends to increase nearly linearly (the objects becoming much fainter) for increasing phase angle \citep{carbognani2019}, while there is a rapid brightness increase for phase angle about below $5^{\circ}-6^{\circ}$, the so called ``opposition effect'' 
\citep{bel2000, carbognani2019}. \\
The study of phase - mag curves are fundamental to determine asteroid absolute magnitudes\footnote{The absolute magnitude of a Solar system object is defined as the (lightcurve-averaged) $V$ magnitude reduced to unit distance from the observer and the Sun, when the body is observed at ideal solar opposition (zero phase angle).} and, in order to have also an independent estimate of the geometric albedo $p_V$, we consider the empirical photometric system proposed by \cite{shev1997}. We have chosen this system because the parameters that characterize it have a more direct physical meaning than the more classic $(H, G)$ and $(H, G_1, G_2)$, though the fit of the mag - phase curves with Shevchenko's system provides residuals slightly higher than the most recent and complex $(H, G_1, G_2)$ system \citep{carbognani2019}. Shevchenko's system is expressed by the following Equation:

\begin{equation}
V(\alpha) = V_{lin}(0)-\frac{a}{1+\alpha}+b\alpha 
\label{shev_sys}
\end{equation}

In Eq.~\ref{shev_sys} $V(\alpha)$ is the observed mag in $V$ band, $\alpha$ is the phase angle in degree and $V_{lin}(0)$, $a$ and $b$ are three constants to be determined. The opposition effect corresponds to the difference between a simple extrapolation to zero phase of the linear part of the mag - phase curve (described by the $b$ parameter), and the value that is actually observed and is determined by the presence of a brightness surge described by the term including the parameter $a$. So in Shevchenko's system the absolute asteroid's mag is given by $V(0)=V_{lin}(0)-a$. 
In Eq.~\ref{shev_sys} $V_{lin}(0)$ represents therefore the extrapolation to zero phase angle of a purely linear mag - phase relation having angular coefficient $b$. \\

From Table~\ref{t02}, using Shevchenko's photometric system with $R_c$ mag, we get the following least squares parameter values: $R_{lin}(0) = 14.73_{\pm 0.04}$ mag, $a=-0.03_{\pm 0.3}$ and $b=0.034_{\pm 0.001}$ mag/degree.\\
In \cite{purdum2021} we found a more complete Gault's mag - phase curve in $r$ band with data taken between 2020 Apr-Oct. The angular coefficient of the linear part is about $b_p\approx 0.03$ (but it is not computed by the authors), in good agreement with our value. The same thing hold for \cite{devogele2021}. \\
The same slope value $b=0.034_{\pm 0.003}$ is found using our V-band mag values, which are only available from Sep onwards: the slope of the linear part of the mag - phase curve is independent from the filter, $V$ or $R_c$, used. Considering the Gault's mean color index $V-R_c = 0.43_{\pm 0.03}$, the corresponding linear mag at zero phase in $V$ band is $V_{lin}(0) = R_{lin}(0) + (V-R_{c}) = 15.16_{\pm 0.05}$ mag. Unfortunately the $a$ value is poorly determined, i.e. the uncertainty is high, because we have practically no observations in the non-linear part of the mag - phase curve (Fig.~\ref{Gault_HG}). For this reason we cannot determine the Gault's absolute mag, but only its superior limit simply given by the linear mag: $V_{lin}(0) = 15.16_{\pm 0.05}$.

In a previous work \citep{carbognani2019}, we found that the corresponding values of the linear part of the Shevchenko system is in good agreement with a simple linear fit of data not including magnitude values affected by the opposition effect. Interestingly, it has been proposed that the slope value of the linear part can be diagnostic of the geometric albedo with the empirical relationship 
\citep{shev1997, bel2000}:

\begin{equation}
b = 0.013_{\pm 0.002} - 0.0104_{\pm 0.0008} \left( \ln {p_V} \right)
\label{shev_albedo}
\end{equation}

In our case we have observed the linear part of the Gault mag - phase curve (see Fig.~\ref{Gault_HG}), and the value of the angular coefficient of the best linear fit is $b=0.034_{\pm 0.002}$ in mag/degree unit. This value is practically coincident with the $b$ value given with the complete best fit of the mag - phase function with Eq.~\ref{shev_sys}, as expected.\\
Our $b$ value is in the range of S-type asteroids, approx between 0.025 and 0.035 \citep{bel2000}. Reversing Eq.~\ref{shev_albedo}, we found $p_V = 0.13_{\pm 0.04}$, a value lower than the average for S-type asteroids, i.e. $p_V = 0.18_{\pm 0.06}$ \citep{wisniewski97}, but still perfectly acceptable. \\
If we know the geometric albedo $p_V$ and the absolute magnitude $V(0)$ we can estimate the effective diameter $D_e$ of an asteroid using the following relation \citep{harrisharris97}: 

\begin{equation}
D_e = \frac{1329}{\sqrt p_V} 10^{-0.2 V(0)} 
\label{DPH}
\end{equation}

In Eq.~\ref{DPH}, $V(0)$ is the absolute magnitude and $p_V$ the geometric albedo in $V$ band. Using the $V_{lin}(0)$ value instead of $V(0)$, we can get the inferior value for Gault's effective diameter.
From Eq.~\ref{DPH} with $p_V \approx 0.13$ and $V(0)\approx 15.2$, we get $D_e \approx 3.4$ km. \\
Considering that there is a non-monotonic dependence between the mean $a$ value, that describe the opposition effect, and the geometric albedo (to $p_V\approx 0.4$ correspond $a\approx 0.3$ mag, see
 \cite{bel2000}), Gault's absolute mag can be estimated as $V(0)\approx +15.2-0.3 \approx +14.9 $. This value is in good agreement with the estimate of absolute mag made by \cite{luu2021} which finds $V(0) = 15.0$ and \cite{devogele2021} with $V(0) = 14.81_{\pm 0.04}$. From Eq.~\ref{DPH}, with $p_V \approx 0.13$ and $V(0)\approx 14.9$, the upper limit for the diameter is about 4 km. \\
Gault's absolute mag reported in the JPL Small-Body Database Browser\footnote{https://ssd.jpl.nasa.gov/}, obtained from astrometric observations, is about +14.3 and this implies a diameter between 2.6 and 8.2 km (geometric albedo from 0.5 to 0.05)\footnote{https://cgi.minorplanetcenter.net/iau/lists/Sizes.html}. Our result suggest that Gault may be in the lower part of this range. \\

\begin{table}
\centering
\caption{Gault's mag - phase data plotted in Fig.~\ref{Gault_HG}. Pre-opposition values are with negative phase angle. The average mags in July was obtained by comparing the luminosity of Gault with that of solar-type stars in the field of view (CMC-15 catalog). From September onwards the mags were obtained using standard calibration with Landolt fields.}
\label{t02}

\begin{tabular}{lcccc}
\hline
Date & Mean UT & Phase ($^{\circ}$) & Mean mag ($R_c$) & Reduced mag ($R_c$)\\

\hline
Jul 21 & 01:20 & -27.45 & $18.36 \pm 0.03$ & $15.67 \pm 0.03$\\
Jul 22 & 02:00 & -27.29 & $18.36 \pm 0.03$ & $15.68 \pm 0.03$\\
Jul 24 & 01:30 & -26.98 & $18.37 \pm 0.04$ & $15.71 \pm 0.04$\\
Jul 27 & 01:30 & -26.26 & $18.15 \pm 0.02$ & $15.53 \pm 0.02$\\
Sep 16 & 22:00 & -06.10 & $17.24 \pm 0.02$ & $14.96 \pm 0.02$\\
Oct 13 & 20:30 & +08.83 & $17.51 \pm 0.01$ & $15.02 \pm 0.01$\\
Oct 16 & 21:50 & +10.30 & $17.64 \pm 0.03$ & $15.11 \pm 0.03$\\
Oct 20 & 21:45 & +12.12 & $17.69 \pm 0.07$ & $15.10 \pm 0.07$\\
Nov 07 & 19:45 & +18.51 & $18.27 \pm 0.01$ & $15.37 \pm 0.01$\\

\hline
\end{tabular}
\end{table}

\section{Summary and conclusions}
We have observed the active asteroid (6478) Gault from Jul to Nov 2020. Photometric observations were not easy due to the low amplitude of the lightcurve and the low declination of the asteroid which limited the time available for observations. \\
Thanks to several dense photometric sessions in $C$ band we have estimated a mean best rotation period $P \approx 2.46_{\pm 0.02}$ h, in agreement with that found by \cite{luu2021}, \cite{purdum2021} and \cite{devogele2021}. In longer sessions of good quality we found a secondary period of about 3.1 h, a value very near the period found in Paper I, i.e. $3.34_{\pm 0.02}$ h. This secondary period, equal to about $5P/4$, appear as an alias of the best period. Changes in lightcurves shape from Aug-Sep (pre-opposition) to Oct (post-opposition) lead us to hypothesize that Gault's surface is a bit irregular and with concavity. Assuming that the asteroid is at the limit of the spin-barrier, the mean density is $\rho\approx 1.80_{\pm 0.03}~\textrm{g}/\textrm{cm}^3$, with an estimated porosity of about 50\%.\\
Gault's appearance was always star-like, no form of detectable cometary activity was seen during our observations. We have determined the average color indices of the asteroid surface ($B-V = +0.84_{\pm 0.04}$, $V-R_{c} = +0.43_{\pm 0.03}$, $B-R_{c} = +1.27_{\pm 0.02}$), finding them compatible with an S-type object. The colors appear change with the rotational phase and there are also indications of an Amplitude-Phase Relationship in $C$ band to be confirmed with more observations.\\
From mean mag values we reconstruct the mag - phase curve in the linear part and from a best fit with Shevchenko's system, we estimated Gault's mean geometric albedo as $p_V = 0.13_{\pm 0.04}$. The albedo, together with an estimate of the absolute mag, allowed us to get Gault's effective diameter as $D_e \approx 3-4$ km. So Gault may be smaller than previously thought based on the absolute mag obtained from astrometric observations.

\section*{acknowledgements}
The authors want to thank Petr Pravec (Ond\v rejov Observatory, Czech Republic), for the independent analysis of Gault's photometric data. Many thanks to the anonymous referee who, with his comments, has considerably increased the quality of the manuscript.

\section*{Data Availability}
The data underlying this article will be shared on reasonable request to the corresponding author.

%-------------------------------------------------------------------

\bsp
\label{lastpage}

\begin{thebibliography}{}

\bibitem[\protect\citeauthoryear{Belskaya et al.}{2000}]{bel2000}
Belskaya, I.\ N., Shevchenko, V.\ G., 2000, Icarus 147, 94-105.

\bibitem[\protect\citeauthoryear{Britt et al.}{2001}]{britt2001}
Britt, T. D. et al., 2001, Lunar and Planetary Science XXXII, 1212

\bibitem[\protect\citeauthoryear{Carbognani \& Buzzoni}{2020}]{carbognani2020}
Carbognani A., Buzzoni A., 2020, Monthly Notices of the Royal Astronomical Society, 493, 70

% Curve di fase degli asteroidi
\bibitem[\protect\citeauthoryear{Carbognani et al.}{2019}]{carbognani2019}
Carbognani A., Cellino A., Caminiti S., 2019, Planetary and Space Science, 169, 15

\bibitem[\protect\citeauthoryear{Dandy et al.}{2003}]{dandy2003}
Dandy, C.~L., Fitzsimmons, A., and Collander-Brown, S.~J., 2003, Icarus, 163, 363

\bibitem[\protect\citeauthoryear{Devogele et al.}{2021}]{devogele2021}
Devog\'ele, M. et al., 2021, Monthly Notices of the Royal Astronomical Society, 505, 245-258

\bibitem[\protect\citeauthoryear{Harris et al.}{2014}]{harris2014}
Harris, A.~W., et al., 2014, Icarus, 235, 55

% Formula per il diametro degli asteroidi
\bibitem[{Harris and Harris(1997)}]{harrisharris97}
Harris, A.\ W., \& Harris, A.\ W., 1997, Icarus 126, 450

\bibitem[\protect\citeauthoryear{Harris et al.}{1981}]{harris1981}
Harris, W.~E., Fitzgerald, M.~P., Reed, B.~C., 1981, PASP, 93, 507

\bibitem[\protect\citeauthoryear{Harris et al.}{1989}]{harris1989}
Harris A.~W., Young J.~W., Bowell E., Martin L.~J., Millis R.~L., Poutanen
M., Zeigler K.~W., 1989, Icarus, 77, 171

\bibitem[\protect\citeauthoryear{Ivanova et al.}{2020}]{ivanova2020}
Ivanova, O. et al., 2020, MNRAS, 496, 2636-2647

\bibitem[\protect\citeauthoryear{Kleyna et al.}{2019}]{kleyna2019}
Jan T. Kleyna et al., 2019, ApJ, 874, L20

\bibitem[\protect\citeauthoryear{Jewitt et al.}{2019}]{jewitt2019}
Jewitt, D.,  Kim, Y., Luu, J., Rajagopa, J., Kotulla, R., Ridgway, S., Liu, W., 2019, ApJ, 876, L19

\bibitem[\protect\citeauthoryear{Jewitt}{2012}]{jewitt2012}
Jewitt, D., 2012, The Astronomical Journal, 143, 66-80

\bibitem[\protect\citeauthoryear{Landolt}{1992}]{landolt1992}	
Landolt, A.~U., 1992, AJ, 104, 340

\bibitem[\protect\citeauthoryear{Luu et al.}{2021}]{luu2021}	
Luu, J.~X., 2021, The Astrophysical Journal Letters, 910, L27

\bibitem[\protect\citeauthoryear{Masiero et al.}{2020}]{masiero2020}
Masiero, J.~R., Mainzer, A.~K., Bauer, J.~M., Cutri, R.~M., Grav, T., Kramer, E., Pittichov\'a, J., Sonnett, S. and
Wright, E.~L., 2020, The Planetary Science Journal, 1, 5

\bibitem[\protect\citeauthoryear{Muinonen et al.}{2010}]{muinonen2010}
Muinonen, K., Belskaya, I.\ N., Cellino, A., Delb\'o, M., Levasseur-Regourd, A.\ C., Penttil\"a, A., Tedesco, E.\ F., 2010, Icarus 209, 542-555

\bibitem[\protect\citeauthoryear{P\'al et al.}{2020}]{Pal2020}
P\'al, A. et al., 2020, ApJS 247, 26

\bibitem[\protect\citeauthoryear{Pravec et al.}{2002}]{pravec2002}
Pravec, P., Ku\v{s}nir\'ak, P., \v{S}arounov\'a, L., Harris, A.~W., 2002, Proc. of Asteroids, 
Comets, Meteors - ACM 2002, 743

\bibitem[\protect\citeauthoryear{Purdum et al.}{2021}]{purdum2021}
Purdum, J.~N. et al., 2021, The Astrophysical Journal Letters, 911, L35

\bibitem[\protect\citeauthoryear{Sanchez et al.}{2019}]{sanchez2019}
Sanchez, J.~A., Reddy, V., Thirouin, A., Wright, E.~L., Linder, T.~R., Kareta, T., Sharkey, B., 2019, The Astrophysical Journal Letters, 881, L6

\bibitem[\protect\citeauthoryear{Shevchenko}{1997}]{shev1997}
Shevchenko, V.\ G., 1997, Solar System Research 31, 219-224.

\bibitem[\protect\citeauthoryear{Smith \& Denneau}{2019}]{smith2019}
Smith, K.~W., \& Denneau, L. 2019, Telegram n. 4594, Central Bureau for Electronic Telegrams

\bibitem[\protect\citeauthoryear{Schwarzenberg-Czerny}{1996}]{schwa1996}
Schwarzenberg-Czerny A., 1996, ApJ, 460, L107

\bibitem[\protect\citeauthoryear{Tanga et al.}{2015}]{tanga2015}
Tanga, P. et al., 2015, Monthly Notices of the Royal Astronomical Society, 448, 3382-3390

\bibitem[\protect\citeauthoryear{Warner}{2009}]{warner2009}
Warner, B.~D., 2009. MPO Software, Canopus. Bdw Publishing. {\sl http://minorplanetobserver.com/}

\bibitem[\protect\citeauthoryear{Wisniewski et al.}{1997}]{wisniewski97}
Wisniewski W. Z., et al., 1997, Icarus, 126, 395

\bibitem[\protect\citeauthoryear{Ye, et al.}{2019}]{quanzhi2019} 
Ye, Q., et al., 2019, ApJ, 874, L16

\bibitem[\protect\citeauthoryear{Zappal\'a et al.}{1990}]{zappala90}
Zappal\'a, V.,\ Cellino, A., Barucci, A. M., Fulchignoni, M., Lupishko D. F., 1990, Astronomy \& Astrophysics 231, 548-560.

\end{thebibliography}
\end{document}